\documentclass[pra,aps,preprint,byrevtex,showpacs]{revtex4}

\usepackage{epsfig}
\usepackage{amsmath}

\newcommand {\ket}[2] {| #1 \rangle_{#2}}
\newcommand {\eqn}[1] {Eq.~(\ref{#1})}
\newcommand {\half} {\frac{1}{2}}
\newcommand {\fig}[1] {Fig.~\ref{#1}}
\newcommand {\figwidth} {80mm}
\newcommand {\Ref}[1] {Ref.~\cite{#1}}

\newcommand {\bout} {\hat{b}_{\mathrm{out}}}
\newcommand {\ain} {\hat{a}_{\mathrm{in}}}

\newcommand {\gain} {g}

\newcommand {\vac} {\hat{v}}
\newcommand {\bmode} {\hat{b}}
\newcommand {\Xplus} {\hat{X}^+}
\newcommand {\Xminus} {\hat{X}^-}

\newcommand {\etal} {\emph{et~al.}}

\begin{document}

\title{Tailoring teleportation to the quantum alphabet}

\author{P.~T.~Cochrane}
\email{cochrane@physics.uq.edu.au}

\author{T.~C.~Ralph}

\affiliation{Department of Physics, The University of Queensland, 
St.~Lucia, Brisbane, Queensland 4072, Australia}

\date{\today}

\begin{abstract}
We introduce a refinement of the standard continuous variable
teleportation measurement and displacement strategies.  This refinement
makes use of prior knowledge about the target state and the
partial information carried by the classical channel when entanglement
is non-maximal.  This gives an improvement in the output quality of
the protocol.  The strategies we introduce could be used in current
continuous variable teleportation experiments.
\end{abstract}

\pacs{03.67.Hk}

\maketitle

\section{Introduction}

Quantum teleportation has become a cornerstone of quantum information
theory since its conception by Bennett~\emph{et al.} in
1993~\cite{Bennett:1993:1}.  It is a useful quantum information
processing task both in itself, and as part of other tasks such as
quantum gate implementation~\cite{Gottesman:1999:2,Knill:2001:1}.  In
particular, optical implementations of
teleportation~\cite{Bouwmeester:1997:1,Furusawa:1998:1,Boschi:1998:1,Bowen:2002:1}
may be useful in current linear optical quantum computing
proposals~\cite{Knill:2001:1}.

Quantum teleportation is a process whereby the state of a quantum
system can be communicated between two (possibly very distant) parties
with prior shared entanglement, joint local quantum measurements,
local unitary transformations and classical communication.  In the
standard scheme, the two parties are called Alice and Bob, and are
sender and receiver respectively.  Victor (the verifier) gives Alice a
quantum system (the target) in a state known only to him.  Alice makes
joint quantum measurements on the target state and her part of the
entanglement resource shared with Bob.  The results of these
measurements she shares with Bob via a classical communication
channel.  This information tells Bob the local unitary transformations
he must perform on his part of the entanglement resource to faithfully
reproduce the target at his location.  Victor then compares the output
state at Bob's location with the target state by calculating the
overlap between the two.  In its simplest form this is just the inner
product of the two states and is in general known as the
\emph{fidelity}.

In ideal teleportation the resource is maximally entangled.  As a
result the classical channel carries no information about the target
state.  Also the alphabet of input states is assumed to be an unbiased
distribution over the same dimensions as the entanglement.  Examples
of this include: the standard discrete protocol where qubits are both
the target and entanglement resource~\cite{Bennett:1993:1}; and the
original continuous variable protocol where the target is a flat,
infinite dimensional distribution and the entanglement is idealised
EPR states\cite{Vaidman:1994:1}.  However, one may consider situations
in which the entanglement is non-maximal and the alphabet of states is
not evenly distributed.  Additional information is now available prior
to teleportation, from the restricted alphabet, and dynamically from
the partial target information now carried by the classical channel.
How should one then tailor the protocol so as to make best use of this
additional information?  We address this question in this paper.

The situation arises naturally in practical implementations of
continuous variable
teleportation~\cite{Braunstein:1998:2,Ralph:1998:1,Furusawa:1998:1}.
The entanglement resource most commonly used in continuous variable
teleportation is the two-mode squeezed vacuum.  It is not perfectly
entangled, since this would require infinite energy.  On the other
hand an even distribution of target states is also unphysical.  We are
motivated to find ways in which to make maximum use of the resource
given this situation.  In this paper we outline a general strategy and
then describe a simple refinement of the standard continuous variable
teleportation protocol which gives an improved output quality for a
reduced alphabet of possible input states.  It has the advantage that
it may be implemented with currently available technology.

Consider the situation of teleporting a coherent state.  The state
amplitudes will have an upper bound, and the probability of Victor
preparing a state with a certain amplitude might be known.  Let us
consider three variations on this theme:
\begin{description}
\item [Two-dimensional Gaussian.] The classical limit used in
\Ref{Furusawa:1998:1} and derived by Braunstein, Fuchs and
Kimble~\cite{Braunstein:2000:1} assumes that Victor produces coherent
states with a symmetric two-dimensional Gaussian probability
distribution, where coherent states of greater amplitude are less
likely to occur than those with amplitude close to zero.  The standard
protocol assumes the width of this distribution is infinite.
Braunstein, Fuchs and Kimble considered how the classical limit
changed for finite width but not how to optimise the protocol as a
function of this width.  Choosing this smaller subset of states should
allow Alice and Bob to improve the fidelity of their teleportation
protocol.
\item [Coherent states on a circle.] Another possibility is that
Victor could produce coherent states of an amplitude known to Alice
and Bob, but of an unknown phase.  If the amplitude of Victor's
prepared coherent states is $\alpha$, then these states will lie on a
circle in phase space of radius $\alpha$, hence the term ``coherent
states on a circle''.  This knowledge reduces the alphabet of possible
output states substantially and should lead to a corresponding
improvement in the fidelity.
\item [Coherent states on a line.] Conversely to coherent states on a
circle, Victor could produce target states of known phase but unknown
amplitude.  These states would lie along a line in phase space and
hence are termed ``coherent states on a line''.  Again, the alphabet
of states is reduced and the fidelity is expected to increase with
respect to the standard protocol.
\end{description}

\section{Tailored displacement strategy}

We now describe a general strategy for tailoring teleportation based
upon maximising the fidelity over Bob's possible displacements in
phase space.  Another technique of fidelity optimisation has been
discussed by Ide~\etal~\cite{Ide:2002:1},
which uses gain tuning to improve the fidelity output.  Our scheme is
similar, however we use the one-shot fidelity of teleporting a
coherent state to find Bob's optimum displacement.  The technique
described here gives very simple relations describing the displacement
Bob must make to achieve the best possible fidelity given the level of
squeezing, Alice's measurement results, and the known properties of
the target state.  Using the transfer operator technique of
Hofmann~\etal~\cite{Hofmann:2000:5}, the one-shot fidelity for
teleportation of a coherent state $\ket{\alpha}{}$ is~\footnote{The
average fidelity $\bar{F}$ is the one-shot fidelity averaged over all
measurement results $\beta$.}
\begin{equation}
F = e^{-|\alpha-\epsilon|^2} e^{-\lambda^2 |\alpha-\beta|^2} \Bigl|
\exp[ \lambda (\alpha^* - \epsilon^*) (\alpha - \beta) ] \Bigr|^2,
\end{equation}
where $\beta = x_- + i p_+$ is a parameter combining Alice's
measurement results of position-difference $x_-$ and momentum-sum
$p_+$, $\lambda$ is the squeezing parameter and $\epsilon$ is the
displacement to be made by Bob.  The variable $\alpha$ is determined
from Alice's measurement results and the prior knowledge about the
target state.  The value of $\alpha$ is therefore a ``best guess'' of
the target given the information at hand.  

Maximising the fidelity over $\epsilon$ finds the displacement Bob
should make on his mode to give the best reproduction of the target
state at his location.  The value of $\epsilon$ which maximises the
fidelity is
\begin{equation}
\epsilon = (1-\lambda)\alpha + \lambda\beta.
\label{eq:epsilonAtMax}
\end{equation}

This has a simple physical interpretation.  In the limit of low
squeezing, the first term dominates and it is best to use whatever
``best guess'' we can make for $\alpha$.  As the level of squeezing
increases, Alice's measurements ($\beta$) become more relevant and the
best guess has less importance.  In the limit of large squeezing the
first term is negligible in comparison to the second term and we are
effectively performing standard continuous variable teleportation.  

\begin{figure}
\centerline{\epsfig{file=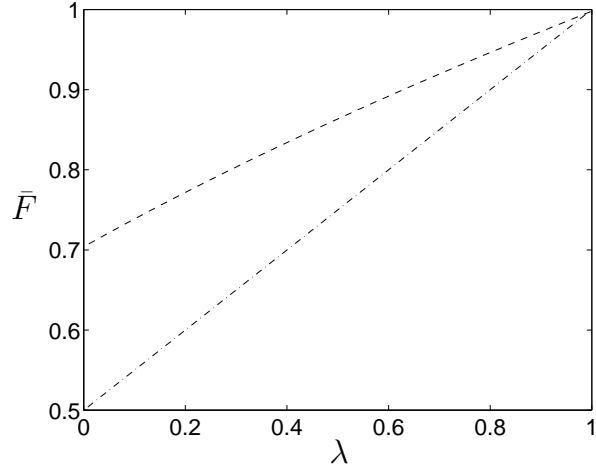, width=\figwidth}}
\caption{Average fidelity $\bar{F}$ versus squeezing parameter
$\lambda$.  The dashed curve is the average fidelity calculated using
the adaptive displacement technique described in the text.  The
dot-dashed curve is the average fidelity produced using standard
continuous variable teleportation.  Using adaptive displacement gives
a large improvement over standard techniques.  The quantities
presented are dimensionless.}
\label{fig:FlambdaDispOnly}
\end{figure}

To illustrate this result, we consider teleportation of states on a
line.  These are simpler to implement experimentally than states on a
circle, since dynamically coordinating the angle of displacement is
more difficult than deciding the size of displacement.  Hence in this
paper we concentrate on states on a line.  We know that the states lie
along the real axis in phase space, therefore $\alpha_y = 0$, and
$\alpha_x$ is determined from information gathered in the
teleportation experiment~\footnote{We use the subscripts $x$ and $y$
to refer to the $x$- and $y$-components respectively of the variables
$\alpha$, $\beta$, and $\epsilon$ in phase space}.  Alice's
measurement result $\beta$ gives this information and we set $\alpha_x
= |\beta|$.  The relations for the $x$- and $y$-components of the
displacement Bob must make are now
\begin{equation}
\epsilon_x = (1-\lambda)|\beta| + \lambda\beta_x \quad\text{and}\quad
\epsilon_y = \lambda \beta_y.
\label{eq:tailoredDispEpsilons}
\end{equation}
Using this technique results in the dashed curve of
\fig{fig:FlambdaDispOnly}, where we observe a significant increase in
fidelity over the standard protocol (dot-dashed curve).

\section{Tailored measurement and displacement strategy}

The relations of \eqn{eq:tailoredDispEpsilons} tailor only the
displacement made by Bob.  A further improvement can be obtained if
one tailors \emph{both} the measurements made by Alice and Bob's
displacement.  It is easier to perform the calculation in the
Heisenberg picture, hence we continue within this formalism.  Consider
the following situation: Alice and Bob know that they are attempting
to teleport coherent states, and they are very sure of the phase of
the states, however the input amplitude is unknown.  What is the best
strategy Alice and Bob can take given that they know the phase of the
input state and the level of squeezing?  The answer is to tailor
Alice's measurements and Bob's displacement to the known amount of
squeezing.  Bob then merely displaces his component of the
entanglement resource in the known direction by an amount related to
the information sent to him.
\begin{figure}
\centerline{\epsfig{file=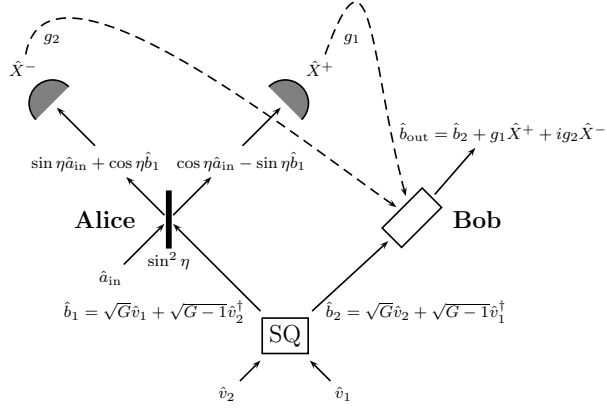, width=\figwidth}}
\caption{Tailored continuous variable teleportation scheme.  The vacua
$\vac_1$ and $\vac_2$ are squeezed in the squeezer SQ, producing the
entangled squeezed beams $\bmode_1$ (which goes to Alice) and
$\bmode_2$ (which goes to Bob).  Alice mixes the target mode $\ain$ on
a beam splitter of reflectivity $\sin^2\eta$ and measures the
quadrature components $\Xplus$ and $\Xminus$.  She modifies these
measurements by the gains $\gain_1$ and $\gain_2$ respectively and
sends the results to Bob via the classical channel, who then displaces
his mode by this amount to obtain a reproduction of the target mode at
his location.}
\label{fig:adapTeleDiag}
\end{figure}
The protocol is described diagrammatically in \fig{fig:adapTeleDiag}
and proceeds as follows: Alice and Bob share one part of a two-mode
squeezed vacuum generated by parametric down conversion of the vacua
$\hat{v}_1$ and $\hat{v}_2$ in the squeezer denoted SQ in the figure.
Once again the phase of the target coherent state is taken to be
zero.  We do not lose generality since it is always possible to rotate
to a frame in which the phase of the target state points along the
real axis in phase space.  Alice mixes her mode $\hat{b}_1$ with that
of the target state $\ain$ on a beam splitter.  The level of mixing is
varied by choosing the beam splitter reflectance $\sin^2\eta$ in a
manner dependent upon the level of squeezing $G$.  Alice makes
measurements of the quadrature observables $\hat{X}^+$ and
$\hat{X}^-$, which are given by
\begin{align}
\hat{X}^+ &= \cos\eta \ain - \sin\eta \hat{b}_1 + \cos\eta
\ain^\dagger - \sin\eta \hat{b}_1^\dagger,
\intertext{and}
\hat{X}^- &= \frac{\sin\eta \ain + \cos\eta \hat{b}_1 - \sin\eta
\ain^\dagger - \cos\eta \hat{b}_1^\dagger}{i}.
\end{align}
Note that for a general mode $\hat{a}$ the quadrature components are
given by
\begin{equation}
\hat{X}^+ = \hat{a} + \hat{a}^\dagger, \quad\text{and}\quad
\hat{X}^- = \frac{\hat{a} - \hat{a}^\dagger}{i}.
\end{equation}
These she modifies by the gain parameters $\gain_1$ and $\gain_2$
respectively before sending this information to Bob via a classical
channel.  The parameters $\gain_1$ and $\gain_2$ are dependent upon
the level of squeezing and the beam splitter reflectance.  Bob uses
this information to displace his mode $\hat{b}_2$ along the real axis
and obtain an approximate reproduction of the initial target state.

The output field from the protocol is
\begin{equation}
\bout = \hat{b}_2 + \gain_1 \hat{X}^+ + i \gain_2 \hat{X}^-,
\end{equation}
from which it is possible show that the quadrature amplitudes of
$\bout$ are
\begin{align}
\hat{X}_{\bout}^+ &= (\sqrt{G} - 2 \gain_1 \sin\eta \sqrt{G-1}) 
\hat{X}_{\hat{v}_2}^+\nonumber\\
&\quad+ (\sqrt{G-1} - 2 g_1 \sin\eta \sqrt{G}) \hat{X}_{\hat{v}_1}^+\nonumber\\
&\quad+ 2 \gain_1 \cos\eta \hat{X}_{\ain}^+,\\
\hat{X}_{\bout}^- &= (\sqrt{G} - 2 \gain_2 \cos\eta \sqrt{G-1}) 
\hat{X}_{\hat{v}_2}^-\nonumber\\
&\quad- (\sqrt{G-1} - 2 \gain_2 \cos\eta \sqrt{G}) \hat{X}_{\hat{v}_1}^-\nonumber\\
&\quad+ 2 \gain_2 \sin\eta \hat{X}_{\ain}^-.
\end{align}
Note that normalisation factors have been absorbed into the gains
$\gain_1$ and $\gain_2$.  This means that at unit gain, when the
output mode is described by
\begin{equation}
\bout = \hat{b}_2 + \frac{1}{\sqrt{2}}(\hat{X}^+ + i \hat{X}^-),
\end{equation}
the gains are $\gain_1 = \gain_2 = \frac{1}{\sqrt{2}}$ instead of
$\gain_1 = \gain_2 = 1$, as for other conventions.

Assuming our states are uniformly distributed along the line (out to
some large $\alpha$) then unit gain for the real quadrature is the
best strategy (as in standard teleportation).  We can determine
$\gain_1$ from this constraint and so we choose
\begin{equation}
\gain_1 = \frac{1}{2 \cos\eta}.
\end{equation}
This value for $\gain_1$ gives the new amplitude quadrature of the 
output mode as
\begin{align}
\hat{X}_{\bout}^+ &= (\sqrt{G} - \tan\eta \sqrt{G-1}) \hat{X}_{\hat{v}_2}^+\nonumber\\
&\quad+ (\sqrt{G-1} - \tan\eta \sqrt{G}) \hat{X}_{\hat{v}_1}^+ + \hat{X}_{\ain}^+.
\end{align}
Unlike the standard protocol, we know that the average value of the
phase quadrature is zero.  Thus we are free to choose the gain on the
phase quadrature, $\gain_2$, such that it maximises the fidelity.
The amplitude and phase quadrature variances of $\bout$ are
\begin{align}
V^+ &= 2G - 4 \tan\eta \sqrt{G(G-1)} + \tan^2\eta (2G-1)\\
V^- &= 2G - 1 - 8 \gain_2 \cos\eta \sqrt{G(G-1)}\nonumber\\
&\quad+ 4 \gain_2^2 (\cos^2\eta (2G-1) + \sin^2\eta).
\end{align}
These values are then substituted into the average fidelity at unit
gain~\cite{Furusawa:1998:1}
\begin{equation}
\bar{F} = \frac{2}{\sqrt{(V^+ + 1)(V^- + 1)}},
\end{equation}
which we now maximise over $\gain_2$.  Maximising the fidelity is
equivalent to minimising the phase quadrature variance $V^-$ over the
same variable.  Performing this minimisation gives the new value of
$\gain_2$
\begin{equation}
\gain_2 = \frac{\cos\eta \sqrt{G(G-1)}}{\cos^2\eta (2G-1) -
\sin^2\eta},
\end{equation}
and the protocol is tailored for states on a line.

Let us consider various limits of the protocol, and measurement and
displacement strategies at those limits.  When there is no squeezing,
one should just measure the amplitude of the incoming state since its
phase is known.  This situation is represented in our protocol by
using a completely transmissive beam splitter and ignoring the
$\hat{X}^-$ measurement.  The parameters in this situation are
therefore $\eta = 0$ (completely transmissive beam splitter), $G = 1$
(no squeezing), and $\gain_2 = 0$ (ignoring all information measured
in the $\hat{X}^-$ quadrature).  This situation gives an amplitude
quadrature variance of $V^+ = 2$ and a phase quadrature variance of
$V^- = 1$, and hence a fidelity of $\bar{F} = \sqrt{\frac{2}{3}}$.
Performing standard teleportation at unit gain with no squeezing gives
a fidelity of $\bar{F} =
\frac{1}{2}$~\cite{Braunstein:1998:2,Braunstein:2000:1}.  One can
therefore see that our protocol gives a good improvement over standard
techniques.  If we choose to teleport using a 50:50 beam splitter we
recover the result for only tailoring the displacement.  With no
squeezing, again the best thing to do is ignore the phase quadrature.
This gives the parameter values $\eta = \frac{\pi}{4}$ (50:50 beam
splitter), $G = 1$ and $\gain_2 = 0$.  However, since we are mixing in
half of the unsqueezed vacuum, we introduce an extra noise component,
increasing the amplitude quadrature variance to $V^+ = 3$ with the
phase quadrature variance being the same at $V^- = 1$; now the
fidelity is $\bar{F} = \frac{1}{\sqrt{2}}$.  For large amounts of
squeezing $G \gg 1$, and it is best to use a 50:50 beam splitter and
perform standard teleportation.  In this limit the quadrature
variances become $V^+ = 1$ and $V^- = 1$ respectively, and the
fidelity tends to unity.

\begin{figure}
\centerline{\epsfig{file=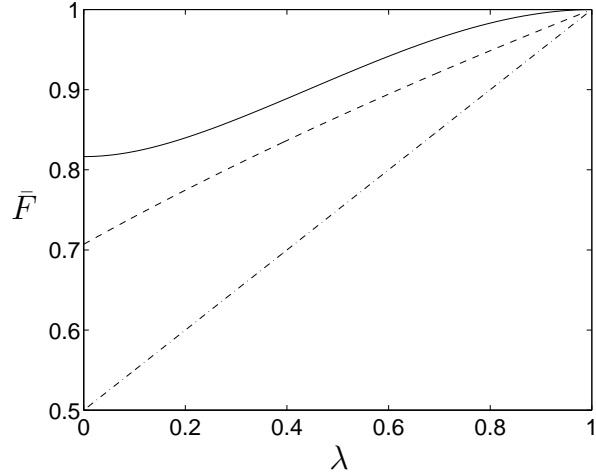, width=\figwidth}}
\caption{Average fidelity $\bar{F}$ as a function of squeezing
parameter $\lambda$ for: tailored measurement and displacement
(solid); tailored displacement (dashed); standard scheme
(dot-dashed).  The quantities presented are dimensionless.}
\label{fig:fidelityLambdaAll}
\end{figure}

In \fig{fig:fidelityLambdaAll} we show these limits graphically and
the trends of three teleportation protocols as a function of squeezing
parameter $\lambda = \sqrt{\frac{G-1}{G}}$.  The solid line represents
the average fidelity as a function of squeezing for the tailored
measurement and displacement scheme.  As mentioned above it starts at
$\bar{F} = \sqrt{\frac{2}{3}}$ at no squeezing ($\lambda = 0$) and
tends to unity as the level of squeezing increases.  Note that this is
a marked improvement over the standard protocol as shown by the
dot-dashed curve.  The dashed curve is the fidelity function when
Alice's beam splitter is set at 50:50, resulting in no tailored
measurement, but still using tailored displacement.  The fidelity
begins at $\bar{F} = \frac{1}{\sqrt{2}}$ and tends to unity with
increasing squeezing.  This too is a good improvement over the
standard protocol.

\begin{figure}
\centerline{\epsfig{file=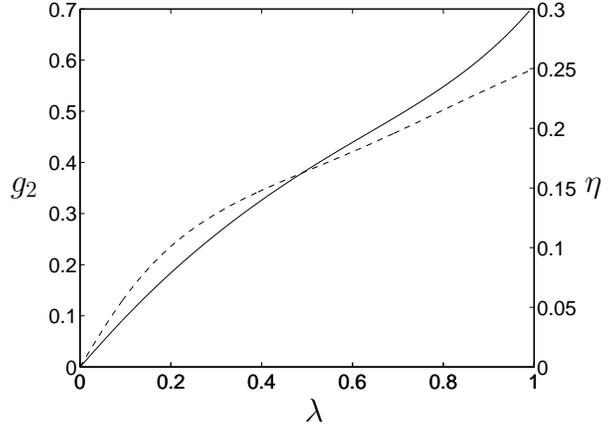, width=\figwidth}}
\caption{Phase quadrature gain $\gain_2$ (solid curve; in
dimensionless units) and beamsplitter parameter $\eta$ (dashed curve;
in units of $\pi$) as a function of squeezing parameter $\lambda$ (in
dimensionless units).  The curves show the values of $\gain_2$ and
$\eta$ one should use to obtain the best fidelity in the tailored
measurement and displacement scheme.}
\label{fig:g2Lambda_etaLambda}
\end{figure}
In order to obtain tailored measurement and displacement average
fidelity as a function of squeezing, one must maximise the fidelity
over both the phase quadrature gain $\gain_2$ and the beam splitter
parameter $\eta$.  The gain and beam splitter parameter values as
functions of squeezing parameter are shown in
\fig{fig:g2Lambda_etaLambda}.  The gain (solid curve) increases
smoothly from zero at no squeezing and tends to $\frac{1}{\sqrt{2}}$
at infinite squeezing ($\lambda = 1$).  The limits are expected since
at no squeezing one does not want to include any information from the
phase quadrature measurement, and hence the gain should be zero.  The
large squeezing limit also makes sense since for large squeezing the
teleporter should be at unit gain, which corresponds to a $\gain_2$
value of $\frac{1}{\sqrt{2}}$.  The beam splitter parameter (dashed
curve) begins at zero at no squeezing and increases smoothly to
$\frac{\pi}{4}$ (note that $\eta$ is given in units of $\pi$ in
\fig{fig:g2Lambda_etaLambda}).  Again, this is sensible behaviour: at
no squeezing one should just measure the target $\ain$ without mixing
in any of the squeezed beam $\hat{b}_1$.  To do this one should have a
completely transmissive beam splitter, which is when $\eta = 0$.  At
infinite squeezing one should equally mix the target and Alice's half
of the entanglement resource.  So, one should use a 50:50 beam
splitter, which corresponds to a beam splitter parameter value of
$\eta = \frac{\pi}{4}$.

The tailored displacement only strategy curve of
\fig{fig:FlambdaDispOnly} calculated from \eqn{eq:epsilonAtMax} is
identical to the equivalent curve in \fig{fig:fidelityLambdaAll}
showing the consistency of the two approaches.  It also turns out, for
$\alpha$ sufficiently large, that using the tailored displacement
scheme to teleport coherent states ``on a circle'' gives the same
fidelity versus squeezing parameter relationship as that found for
teleporting coherent states on a line.  Bob's displacement in this
instance has $x$- and $y$-components
\begin{align}
\epsilon_x &= (1-\lambda) |\alpha| \cos[\arg(\beta)] + \lambda \beta_x\\
\epsilon_y &= (1-\lambda) |\alpha| \sin[\arg(\beta)] + \lambda \beta_y,
\end{align}
where $|\alpha| \cos[\arg(\beta)]$ and $|\alpha| \sin[\arg(\beta)]$
are the best guesses for $\alpha_x$ and $\alpha_y$ respectively, and
$\arg(\beta) = \tan^{-1}(\beta_y/\beta_x)$.  This result is supported
by the paper of Ide~\etal~\cite{Ide:2002:1}
where they too discussed the optimal teleportation of coherent states
of known amplitude, but unknown phase and showed an average fidelity
versus squeezing parameter relationship very similar to that shown in
\fig{fig:FlambdaDispOnly} of this paper.  That states on a line and
states on a circle have the same fidelity relationship indicates that
the two situations are interchangeable; the trends from one can be
used to give the results for the other.  Further improvement of
teleportation for states on the circle would require the use of
adaptive phase measurements~\cite{Wiseman:1995:3}.

\section{Two-dimensional distributions in phase space}

We now adapt the tailored displacement scheme in the Heisenberg
picture to the situation of the target state alphabet being a
two-dimensional distribution in phase space.  Let's begin by deriving
the fidelity of teleportation for a variable linear gain $\gain$
applied to both Alice's measurement results, and a target field $\ain$
mixed with Alice's part of the two-mode squeezed vacuum entanglement
resource on a 50:50 beam splitter.  To do this we calculate the
variance of the teleporter output field $\bout$.  For a level of
squeezing $G$, the output field amplitude quadrature $\Xplus_{\bout}$
can be shown to be
\begin{equation}
\Xplus_{\bout} = (\sqrt{G} - \gain \sqrt{G-1}) \Xplus_{\vac_2} +
(\sqrt{G-1} - \gain \sqrt{G}) \Xplus_{\vac_1} - \gain \Xplus_{\ain},
\end{equation}
where $\vac_1$ and $\vac_2$ are the vacua prior to being squeezed in
the parametric down converter.  This is the same situation as in
\fig{fig:adapTeleDiag} where $\gain_1 = \gain_2 = \gain/\sqrt{2}$ and
$\eta = \frac{\pi}{4}$.  It is possible to show that the variance of
this quadrature is
\begin{equation}
V^+ = 2 G - 4 \gain \sqrt{G(G-1)} + 2 \gain^2 G - 1.
\end{equation}
The phase quadrature and its corresponding variance are equal to
$\Xplus_{\bout}$ and $V^+$ respectively.  This is now sufficient
information to calculate the average fidelity, which for a general
gain has the form~\cite{Furusawa:1998:1}
\begin{equation}
\bar{F}(\alpha) = \frac{2}{\sqrt{(V^+ + 1) (V^- + 1)}} \exp\Biggl[ -\frac{2
|1-\gain|^2 |\alpha|^2}{\sqrt{(V^+ + 1) (V^- + 1)}}\Biggr],
\label{eq:fidHeisGen}
\end{equation}
where $\alpha$ is the amplitude of the coherent state being
teleported, and $V^-$ is the phase quadrature variance.

To make use of the knowledge of the target state alphabet, we weight
this fidelity by the probability of Victor preparing a given target
state $\ket{\alpha}{}$.  A simple case of this probability is a
two-dimensional Gaussian distribution centred at the origin in phase
space.  This form of the distribution will be used in the following
discussion since it was used by Braunstein, Fuchs and
Kimble~\cite{Braunstein:2000:1}, hence their results can be compared
with those presented here.

The probability of Victor preparing a given state $\ket{\alpha}{} =
\ket{\alpha_x + i \alpha_y}{}$ in phase space is given by
\begin{equation}
P(\alpha) = \frac{1}{2\pi s_x s_y}
\exp\Biggl[-\frac{\alpha_x^2}{2 s_x^2} - \frac{\alpha_y^2}{2 s_y^2}\Biggr],
\label{eq:twoDDist}
\end{equation}
where $s_x$ and $s_y$ are the standard deviations of the Gaussian in
the $x$- and $y$-directions respectively.  An overall figure of merit
for the teleportation protocol is the optimised average fidelity
defined by
\begin{equation}
\bar{\mathcal{F}} = \int \bar{F}(\alpha) P(\alpha) d\alpha,
\label{eq:avAvFid}
\end{equation}
where the integral is taken over all possible values of $\alpha$.
Notice that the optimised average fidelity is an implicit function of
the gain $\gain$, and the amount of squeezing $G$.  In the examples
that follow, this optimised average fidelity is numerically maximised
over the gain to find the relationship between the average fidelity
and the level of squeezing, $\lambda$, for the given alphabet of
target states.

Let's now look at two examples of probability distributions in phase
space.  We shall initially analyse a symmetric Gaussian with very
large standard deviation, the reason being that this example
corresponds well with standard continuous variable teleportation.  We
will then investigate a very narrow symmetric distribution and compare
the no squeezing limit with the classical level derived by Braunstein,
Fuchs and Kimble.

\begin{figure}
\centerline{\epsfig{file=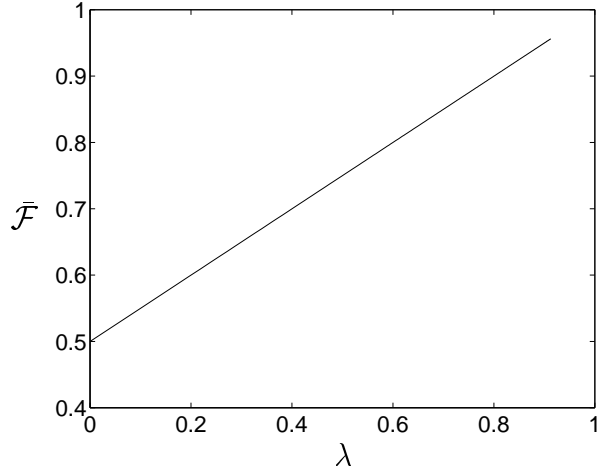, width=\figwidth}}
\caption{Optimised average fidelity maximised over the gain
$\bar{\mathcal{F}}$ as a function of squeezing parameter $\lambda$ for
a very broad symmetric two-dimensional Gaussian distribution of target
states in the complex plane.  The standard deviation of the Gaussian
is $s = 100$.  The relationship is identical to that of the dot-dashed
curve in \fig{fig:fidelityLambdaAll} which is the standard continuous
variable teleportation result.  This is expected, since such a broad
Gaussian is a good approximation of a flat distribution in phase
space.  The gain at the average fidelity maximum is $\gain = 1$ as
expected for standard continuous variable teleportation.  The
quantities presented are dimensionless.}
\label{fig:FlambdaSymWide}
\end{figure}

In the first example, the distribution is symmetric with standard
deviation $s = s_x = s_y = 100$.  Such a distribution is a very good
approximation of the situation in standard continuous variable
teleportation, where it is assumed that the alphabet of target states
is a flat distribution over all phase space.  Calculating the
optimised average fidelity $\bar{\mathcal{F}}$ maximised over the gain
as a function of squeezing parameter $\lambda = \sqrt{\frac{G-1}{G}}$
one obtains the trend in \fig{fig:FlambdaSymWide}.
The optimised average fidelity increases linearly from
$\bar{\mathcal{F}} = \half$ at no squeezing to $\bar{\mathcal{F}} = 1$
at infinite squeezing.  This is the \emph{same curve} as the
dot-dashed curve in \fig{fig:fidelityLambdaAll}.  This result is
expected since, as mentioned above, a very broad Gaussian distribution
is a good approximation to the perfectly flat distribution of target
states assumed in the standard protocol.

\begin{figure}
\centerline{\epsfig{file=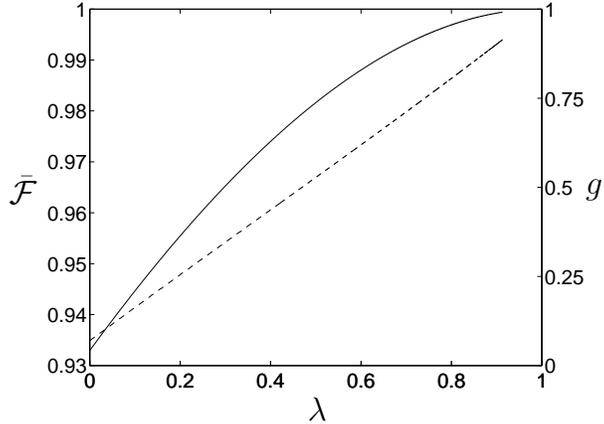, width=\figwidth}}
\caption{Optimised average fidelity maximised over the gain
$\bar{\mathcal{F}}$ (solid curve), and its respective gain $\gain$
(dashed curve), as a function of squeezing parameter $\lambda$ for a
narrow symmetric two-dimensional Gaussian distribution of target
states in the complex plane.  The standard deviation of the
distribution is $s = 0.2$.  The average fidelity at no squeezing
corresponds to the prediction of Braunstein, Fuchs and Kimble for a
distribution of this standard deviation.  The average fidelity then
increases to unity as the level of squeezing increases.  The gain
curve indicates that as squeezing increases the optimal gain will tend
to unity.  The quantities presented are dimensionless.}
\label{fig:FlambdaSym}
\end{figure}

In the second example, the distribution is symmetric with standard
deviation $s = s_x = s_y = 0.2$.  This is a very narrow distribution
and one would expect to find a high optimised average fidelity for all
levels of squeezing since the alphabet of target states does not
deviate far from the vacuum.  Performing tailored displacement
teleportation using this distribution as the alphabet of target states
and maximising the optimised average fidelity over the gain, one
produces the optimised average fidelity versus squeezing parameter
relationship shown in \fig{fig:FlambdaSym}.

Note that the optimised average fidelity at $\lambda = 0$ (i.e.~the
classical level) is very much greater than the value of
$\bar{\mathcal{F}} = \half$ normally predicted by standard
teleportation.  Braunstein, Fuchs and Kimble~\cite{Braunstein:2000:1}
derived a relationship between the average fidelity at the classical
limit and the spread of the two-dimensional Gaussian, optimised for
the given distribution.  This relationship is
\begin{equation}
\bar{\mathcal{F}} = \frac{1+\chi}{2+\chi},
\label{eq:fidBFK}
\end{equation}
where $\chi$ is inversely proportional to the square of the standard
deviation of the Gaussian.  In the discussion here $\chi = \frac{1}{2
s^2}$ where $s = s_x = s_y$.  For a symmetric Gaussian of standard
deviation $s = 0.2$, using \eqn{eq:fidBFK} one would expect the
optimised average fidelity at the classical level to be
$\bar{\mathcal{F}} = 0.931$.  At $\lambda = 0$ in \fig{fig:FlambdaSym}
one finds that $\bar{\mathcal{F}} = 0.933$.  A similar level of
agreement exists for all values of the standard deviation $s$.  Thus,
to a good approximation, our results agree with the classical limit of
Braunstein, Fuchs and Kimble.

Overall, one can still make use of the prior knowledge of the
target state alphabet and optimise the protocol over the gain for
nonzero levels of squeezing.  This is what has been done here; the
fidelity increases from the classical level up to unity with
increasing squeezing as shown explicitly in \fig{fig:FlambdaSym}.  The
tailored displacement teleportation technique again being useful in
improving continuous variable teleportation.

\section{Summary}

We have introduced a refined measurement and displacement strategy
which makes good use of the properties of prior knowledge about the
target state and non-maximal entanglement.  This refinement is
tailored to the given experimental situation and shown to give a great
improvement on the output quality of continuous variable
teleportation.  The two techniques of calculating the tailored
displacement strategy gave identical results, and some physical
insight into how this strategy works. 

We also analysed symmetric two-dimensional Gaussian distributions of
coherent states as an alphabet of target states.  We showed agreement
with the results of Braunstein, Fuchs and Kimble, and extended their
work by including squeezing in the model.

The strategy described here is generally applicable to all
teleportation schemes involving physically limited resources.  A major
advantage of this scheme is that it is able to be implemented with
current continuous variable teleportation technology since it only
requires linear gain on the measurement results.

\begin{acknowledgments}
PTC acknowledges the financial support of the Centre for Laser
Science, the University of Queensland and the Australian Research
Council.  The authors thank G.~J.~Milburn for helpful discussions and
insights in the production of this work.
\end{acknowledgments}

\end{document}